\documentclass{emulateapj}

\usepackage{amsmath}

\newcommand{\tmin}{\ensuremath{t_{\rm min}}}
\newcommand{\tspin}{\ensuremath{\tau_{\rm spin}}}

\newcommand{\UAB}{\ensuremath{U_{\rm AB}}}
\newcommand{\UD}{\ensuremath{U_{\rm D}}}
\newcommand{\UE}{\ensuremath{U_{\rm E}}}

\newcommand{\PDF}{\ensuremath{F}}

\newcommand{\chisq}{\ensuremath{\chi^{2}}}
\newcommand{\chisqr}{\ensuremath{\chi^{2}_{\rm red}}}

\newcommand{\cycnum}{\ensuremath{E}}

\newcommand{\OC}{\ensuremath{O-C}}
\newcommand{\OCmod}{\ensuremath{(\OC)^{\rm mod}}}

\newcommand{\pnull}{\ensuremath{p_{\rm null}}}

\newcommand{\OCstat}{Q}

\newcommand{\soriab}{$\sigma$~Ori~AB}
\newcommand{\sorid}{$\sigma$~Ori~D}
\newcommand{\sorie}{$\sigma$~Ori~E}

\newcommand{\iraf}{\textsc{iraf}}

\newcommand{\eminus}{e$^{-}$}

\begin{document}

%% Title

\title{Discovery of Rotational Braking in the Magnetic Helium-Strong Star Sigma Orionis E}
 \shorttitle{Discovery of Rotational Braking in the Magnetic Helium-Strong Star Sigma Orionis E}

%% Authors

 \author{R. H. D. Townsend}\affil{Department of Astronomy, University
   of Wisconsin-Madison, Sterling Hall, 475 N. Charter Street,
   Madison, WI 53706, USA; townsend@astro.wisc.edu}
 \author{M. E. Oksala}\affil{Bartol Research Institute, Department of
   Physics and Astronomy, University of Delaware, Newark, DE 19716,
   USA} \author{D. H. Cohen}\affil{Department of Physics and
   Astronomy, Swarthmore College, Swarthmore, PA 19081, USA}
 \author{S. P. Owocki}\affil{Bartol Research Institute, Department of
   Physics and Astronomy, University of Delaware, Newark, DE 19716,
   USA} \and \author{A. ud-Doula}\affil{Penn State Worthington
   Scranton, 120 Ridge View Drive, Dunmore, PA 18512, USA}

\shortauthors{Townsend et al.}

%% Abstract

\begin{abstract}
  We present new $U$-band photometry of the magnetic Helium-strong
  star \sorie, obtained over 2004--2009 using the SMARTS 0.9-m
  telescope at Cerro Tololo Inter-American Observatory. When combined
  with historical measurements, these data constrain the evolution of
  the star's 1\fd19 rotation period over the past three decades. We
  are able to rule out a constant period at the $\pnull = 0.05\%$
  level, and instead find that the data are well described ($\pnull =
  99.3\%$) by a period increasing linearly at a rate of 77\,ms per
  year. This corresponds to a characteristic spin-down time of
  1.34\,Myr, in good agreement with theoretical predictions based on
  magnetohydrodynamical simulations of angular momentum loss from
  magnetic massive stars. We therefore conclude that the observations
  are consistent with \sorie\ undergoing rotational braking due to its
  magnetized line-driven wind.
\end{abstract}


%% Introduction

\section{Introduction} \label{sec:intro}

The helium-strong star \sorie\ (HD~37479; B2Vpe; $V=6.66$) has long
been known to harbor a circumstellar magnetosphere in which plasma is
trapped and forced into co-rotation by the star's strong ($\sim
10\,{\rm kG}$) dipolar magnetic field \citep[see,
e.g.,][]{LanBor1978,GroHun1982}. This magnetosphere is largely
responsible for the star's distinctive eclipse-like dimmings, which
occur when plasma clouds transit across the stellar disk twice every
1\fd19 rotation cycle \citep{TowOwo2005}. Some fraction of the star's
photometric variations likely also arise from its photospheric
abundance inhomogeneities, as in other chemically peculiar stars
\citep[e.g,][]{Mik2009}; but for \sorie\ the magnetospheric
contribution to the variations is dominant \citep{Tow2008}.

This paper presents new $U$-band photometry of the star's primary
light minimum\footnote{The term `primary' stems from early
  mis-identifications of the star as an eclipsing binary system
  \citep[e.g.,][]{Hes1976}; here, it simply indicates the deeper of
  the star's two light minima.}, obtained over four seasons spanning
2004--2009 using the SMARTS 0.9-m telescope at Cerro Tololo
Inter-American Observatory (CTIO). When combined with historical
measurements by \citet{Hes1977}, these new data allow a precise
measurement of the star's rotation period, and its evolution, over the
past three decades.

A description of the observations, both archival and new, is provided
in the following section. In \S\ref{sec:analysis}, we discuss a
procedure for accurately measuring the times \tmin\ of primary light
minimum, and then use these measurements to construct a standard
observed-minus-corrected (\OC) diagram for the star, allowing us to
assess the evolution of the star's rotation period. We discuss and
summarize our findings in \S\ref{sec:discuss}.

\section{Observations} \label{sec:obs}

\begin{figure*}[htb]
\epsscale{1}
%\begin{centering}
%\includegraphics{fig_minima}
\plotone{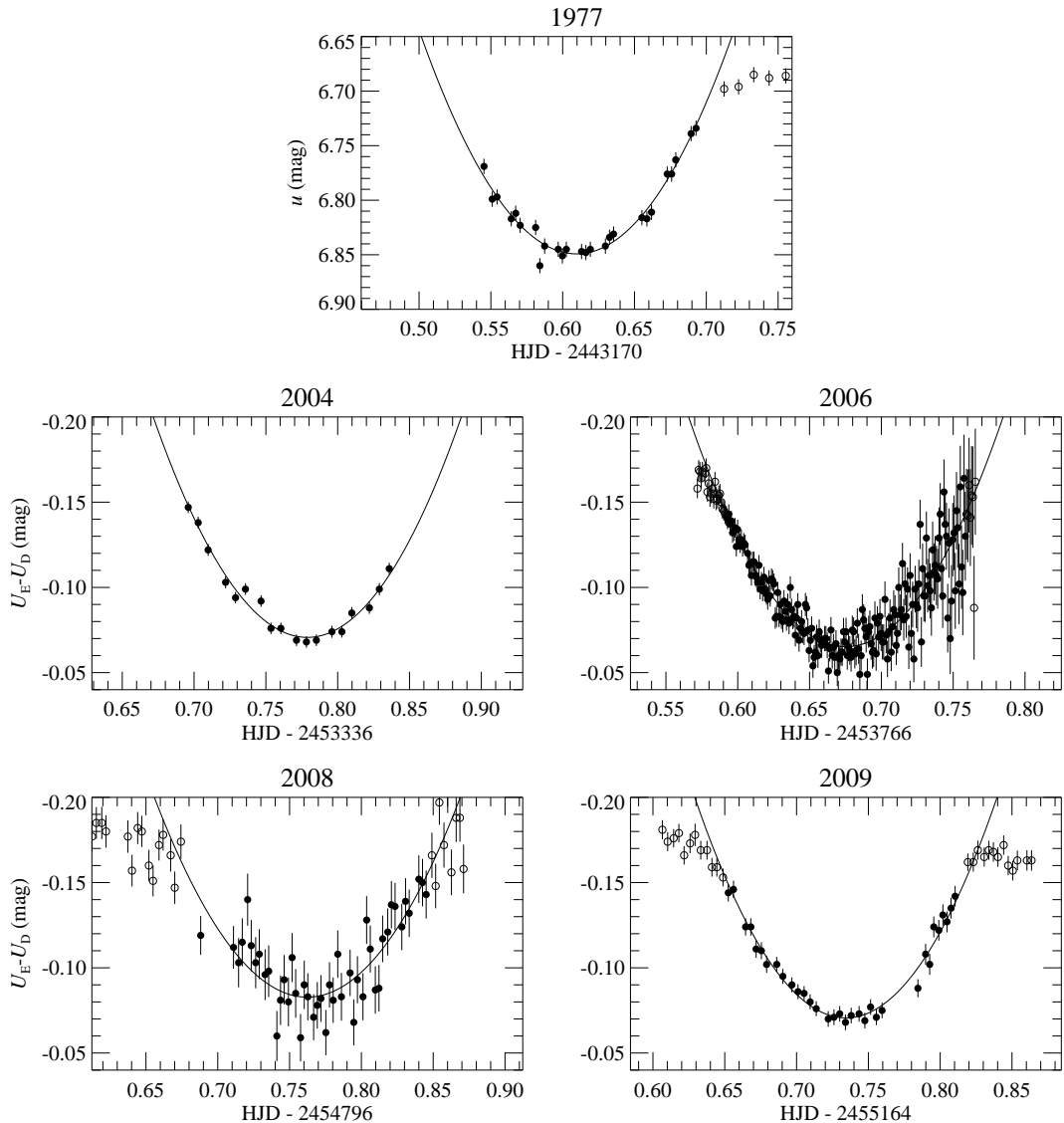}
\caption{Light curves for the primary minima across the five seasons
  (\S\ref{sec:obs}); filled symbols denote those points that
  contribute to the minimum fitting, and open symbols those that do
  not. Vertical bars indicate the estimated measurement errors, and
  the best-fit parabolae (\S\ref{ssec:analysis-min}) are drawn as the
  solid curves.} \label{fig:minima}
%\end{centering}
\end{figure*}

\subsection{1977} \label{ssec:obs-1977}

\citet{Hes1977} observed a primary light minimum of \sorie\ on the
night of 1977 January 26/27, as part of their long-term Str\"omgren
$uvby$ monitoring of the star using the number 1 0.4-m telescope at
CTIO. Their photometric data were kindly provided to us in electronic
form by Prof. C. T. Bolton. No error estimates were supplied, so a
measurement error $\Delta u = 7\,{\rm mmag}$ is assumed for all data
points --- the value quoted by \citet{Hes1976} as an upper limit on
their photometric uncertainties. We do not correct for the apparent
season-to-season brightening noted by \citet{Hes1977}\footnote{This
  brightening likely stems from long-term variations in the comparison
  star HR~1861; see \citet{Ols1977}.}, since this has no effect on the
\tmin\ determinations. The $u$-band light curve is plotted in
Fig.~\ref{fig:minima}; the accompanying $vby$ data are not shown,
since they play no direct role in the period determination (however,
see \S\ref{ssec:analysis-sys}).

\subsection{2004} \label{ssec:obs-2004}

We observed a primary light minimum on the night of 2004 November
26/27, during Johnson $UBVRI$ monitoring of \sorie\ using the
Cassegrain-focus Tek 2048 CCD on the SMARTS 0.9-m telescope
\citep[see][for the full dataset]{OksTow2007}. For all $U$-band
exposures (in every case, $420\,{\rm s}$ long), the Tek \#1 filter
(center 3575\,\AA, FWHM 600\,\AA) was used in tandem with the ND3
neutral density filter (7.5\,mag attenuation). CCD frames were reduced
using standard \iraf\ tasks for bias subtraction and flat-fielding,
and cosmic rays were cleaned via Laplacian edge detection
\citep{vanD2001}.  The \emph{phot} task of the DAOPHOT package, with
an aperture radius of 10 pixels ($=3.96''$) and a sky annulus radius
of 15--20 pixels, was used to perform synthetic aperture
photometry. Measurement errors $\Delta U$ were estimated as the sum in
quadrature of the photometric noise reported by \emph{phot}, the CCD
read noise, and the atmospheric scintillation noise described by the
expression on p.~141 of \citet{Bir2006}.

As a comparison star, we adopt the nearby ($30''$) \sorid\ (HD~37468D;
B2V; $V=6.62$) due to its color and brightness similarity; the
$\UE-\UD$ differential light curve is plotted
Fig.~\ref{fig:minima}. To make an independent check on the constancy
of \sorid, we compare its $U$-band data against \soriab\ (HD~37468;
O9V+B0.5V; $V=3.80$). Over the night, we find a mean difference
$\overline{\UD-\UAB} = 3.240\,{\rm mag}$ and a standard deviation
$\sigma(\UD-\UAB) = 2.8\,{\rm mmag}$; the latter is consistent with the
mean measurement error $\overline{\Delta(\UD-\UAB)} = 2.6\,{\rm
  mmag}$.

\subsection{2006} \label{ssec:obs-2006}

We observed a primary light minimum on the night of 2006 January
30/31. The one modification to the 0.9-m instrumental setup was the
removal of the ND3 neutral density filter from the light path, thereby
shortening exposure times to $2\,{\rm s}$ (with the idea of obtaining
more measurements during the night). In hindsight, this was a
counterproductive move. Although the proximity of \sorid\ and \sorie\
mean that shutter corrections are unimportant, the exposures are
strongly affected by scintillation noise, especially toward the end of
the night when the airmass becomes large. Moreover, \soriab\ is
saturated in all CCD frames and cannot be used as a check star.

The observations were reduced and analyzed as before; the $\UD-\UE$
light curve is plotted in Fig.~\ref{fig:minima}.

\subsection{2008} \label{ssec:obs-2008}

We observed a primary light minimum on the night of 2008 November
25/26. To obtain more-reasonable ($22-65\,{\rm s}$) exposure times
than in the 2006 season, we reintroduced a neutral density filter
(ND2; $5.0\,{\rm mag}$). Changes to the default telescope
configuration led to the inadvertent substitution of the Tek \#2
$U$-band filter (center 3570\,\AA, FWHM 660\,\AA) in place of the Tek
\#1 filter used previously, but this should have negligible impact on
our results. Due to a further oversight, the CCD gain was adjusted
from the previous setting of 3.1 \eminus/ADU, to 0.6 \eminus/ADU; this
had the unfortunate effect of significantly elevating the readout
noise.

The observations were reduced and analyzed as before; the $\UE-\UD$
light curve is plotted in Fig.~\ref{fig:minima}. Comparison of \sorid\
against \soriab\ reveals a mean $\overline{\UD-\UAB} = 3.236\,{\rm
  mag}$ and a standard deviation $\sigma(\UD-\UAB) = 12\,{\rm
  mmag}$. The latter value is rather larger than the mean measurement
error $\overline{\Delta(\UD-\UAB)} = 9.8\,{\rm mmag}$, but not overly
so.

\subsection{2009} \label{ssec:obs-2009}

We observed a primary light minimum on the night of 2009 November
28/29. We kept the ND2 filter from the 2008 season, but reverted back
to the original Tek \#1 $U$-band filter, and set the CCD gain to 1.5
\eminus/ADU; this led to exposure times between 130 and $250\,{\rm
  s}$.  The observations were reduced and analyzed as before; the
$\UE-\UD$ light curve is plotted in Fig.~\ref{fig:minima}. Comparison
of \sorid\ against \soriab\ reveals a mean $\overline{\UD-\UAB} =
3.224\,{\rm mag}$ and a standard deviation $\sigma(\UD-\UAB)
=4.2\,{\rm mmag}$, the latter consistent with the mean measurement
error $\overline{\Delta(\UD-\UAB)} = 4.0\,{\rm mmag}$.

\section{Analysis} \label{sec:analysis}

\begin{figure*}[htb]
\epsscale{1}
%\begin{centering}
%\includegraphics{fig_pdf}
\plotone{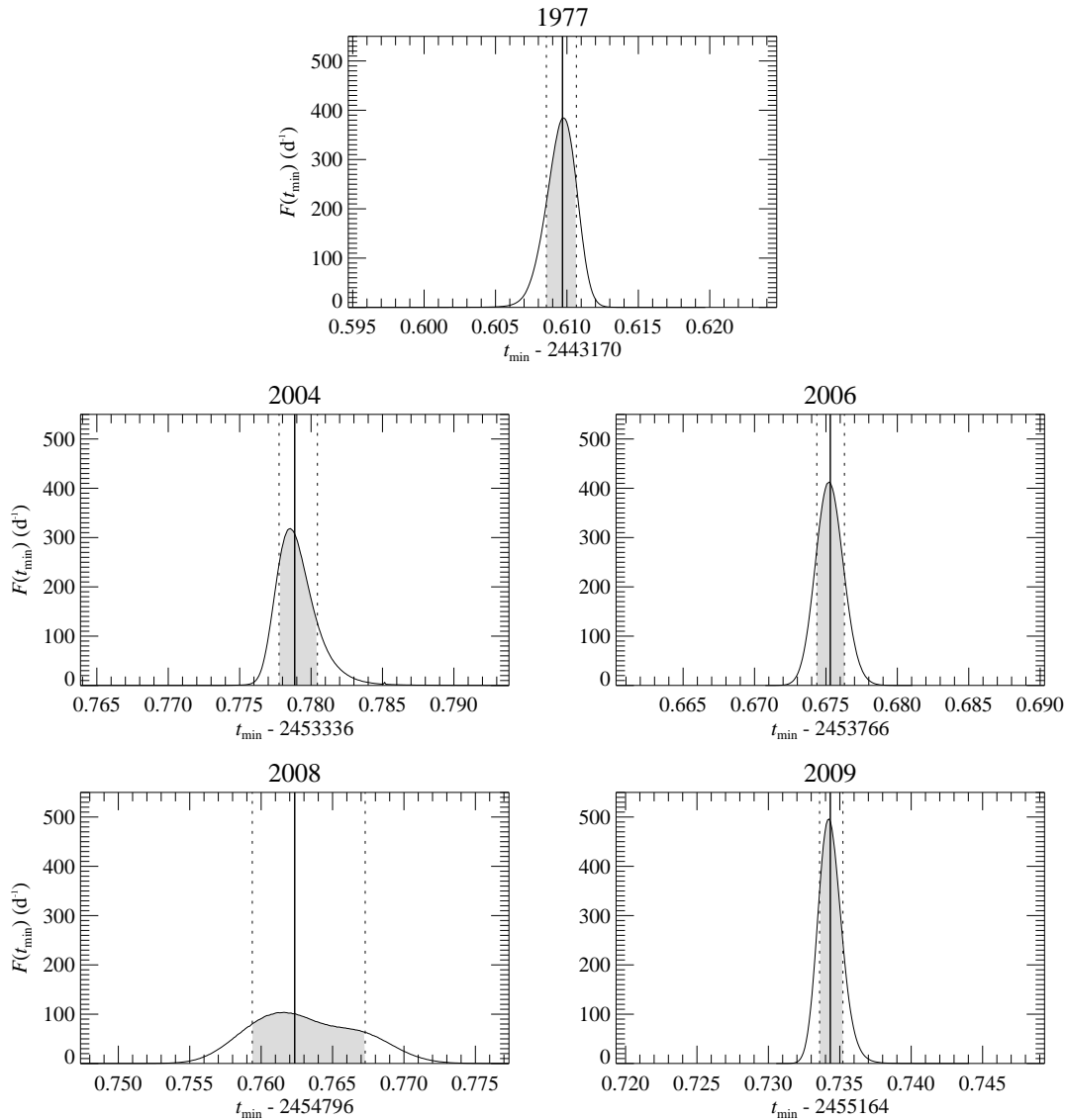}
\caption{Probability distribution functions for the time of primary
  minimum; the thick vertical line indicates the measured \tmin. The
  shaded regions to the left and right of the line each constitute
  34.1\% of the area under the PDF, and define the lower and upper
  error bounds on \tmin\ (shown as dotted vertical
  lines).} \label{fig:pdf}
%\end{centering}
\end{figure*}

\subsection{Minimum Measurements} \label{ssec:analysis-min}

\begin{deluxetable}{lcccc}
\tablecaption{Light-curve minima from parabola fitting\label{tab:minima}}
\tablehead{
Season & Band & \tmin\ (d) & \chisqr & \cycnum \\
}
\tablecolumns{7}
\tabletypesize{\footnotesize}
\startdata
1977 & $u$ &  2443170.6097 $^{+ 0.0010}_{-0.0011}$ & 1.01 & 329 \\ 
2004 & $U$ &  2453336.7789 $^{+ 0.0016}_{-0.0011}$ & 1.56 & 8866 \\ 
2006 & $U$ &  2453766.6753 $^{+0.0010}_{-0.0009}$ & 1.11 & 9227 \\ 
2008 & $U$ &  2454796.7624 $^{+ 0.0049}_{-0.0030}$ & 1.12 & 10092 \\ 
2009 & $U$ &  2455164.7343 $^{+ 0.0009}_{-0.0007}$ & 0.75 & 10401 \\ 
\enddata
\end{deluxetable}

The determination of the time of light minimum (or maximum) in
astronomical objects has received significant attention in the
literature \citep[see][and references therein]{Ste2005}. Historically
the \citet{KweVanW1956} algorithm had long been the standard tool, but
more recently polynomial fitting has emerged as a powerful and robust
approach. Thus we determine the time of light minimum \tmin\ for each
of the five light curves plotted in Fig.~\ref{fig:minima} by an
adaptive parabola fitting procedure:

\begin{enumerate}
\item The time of the dimmest point in the light curve is chosen as
  the initial \tmin.
\item Weighted \chisq\ minimization is used to fit a parabola to those
  points lying within 2 hours of the current \tmin\ (this time
  interval is chosen to exclude the non-eclipse parts of the light curve).
\item A new \tmin\ is chosen by analytically evaluating the minimum of
  the fitted parabola.
\item Steps (ii)-(iii) are repeated until the value of \tmin\ no
  longer changes. (Typically, 3--4 of these iterations are required.)
\end{enumerate}

The parabolic fits are drawn over the light curves in the figure, and
Table~\ref{tab:minima} documents the \tmin\ and reduced \chisq\ values
associated with each fit. Here and throughout, times are expressed as
heliocentric Julian dates (HJD). The tabulated \tmin\ error bounds are
calculated from bootstrap Monte Carlo simulations
\citep{Pre1992}. Specifically, for a given light curve comprising $n$
points, a synthetic light curve is generated by selecting $n$ points
at random with replacement. The parabola fitting procedure is then
used to determine \tmin\ for the synthetic curve. This sequence is
repeated many times to build up a population of synthetic \tmin\
values, from which we derive a probability distribution function
$\PDF(\tmin)$ reflecting a best estimate of the distribution from
which the actual \tmin\ measurement is drawn.

Figure~\ref{fig:pdf} plots the probability distribution functions
(PDFs) associated with the five \tmin\ measurements; each is based on
a population of $10^{7}$ synthetic light curves. The error bounds
quoted in Table~\ref{tab:minima} are the confidence limits relative to
\tmin\ that each enclose 34.1\% of the area under the PDF. Although
this choice mirrors the 1-$\sigma$ limits of a Gaussian distribution,
the PDFs in Fig.~\ref{fig:pdf} underscore that the analysis here does
not presume Gaussian errors. Moreover, while the final error in \tmin\
is insensitive to the {\em estimates} made for the measurement errors
on the individual photometric data, the Monte Carlo simulations used
to derive the PDFs do properly allow the {\em actual} errors of the
noisier data, e.g. in 2008, to produce a larger uncertainty in \tmin.

\subsection{\OC\ Fitting} \label{ssec:analysis-oc}

\begin{deluxetable}{lcccc}
\tablecaption{Fits to the \OC\ data\label{tab:o-c}}
\tablehead{
Fit Type & $b_{1} \times 10^{2}$ (d) & $b_{2} \times 10^{5}$ (d) & $b_{3} \times 10^{9}$ (d) & \pnull\ (\%) \\
}
\tablecolumns{6}
\tabletypesize{\footnotesize}
\startdata
Linear             & 0.38 $^{+ 0.12}_{-0.12}$ & 2.78 $^{+ 0.02}_{-0.01}$ & --                     &  0.05 \\
Quadratic          & 1.00 $^{+ 0.12}_{-0.13}$ & 1.29 $^{+ 0.12}_{-0.10}$ & 1.44 $^{+ 0.10}_{-0.11}$ & 99.3 \\
\enddata
\end{deluxetable}

\begin{figure}[htb!]
\epsscale{1}
%\begin{centering}
%\includegraphics{fig_oc}
\plotone{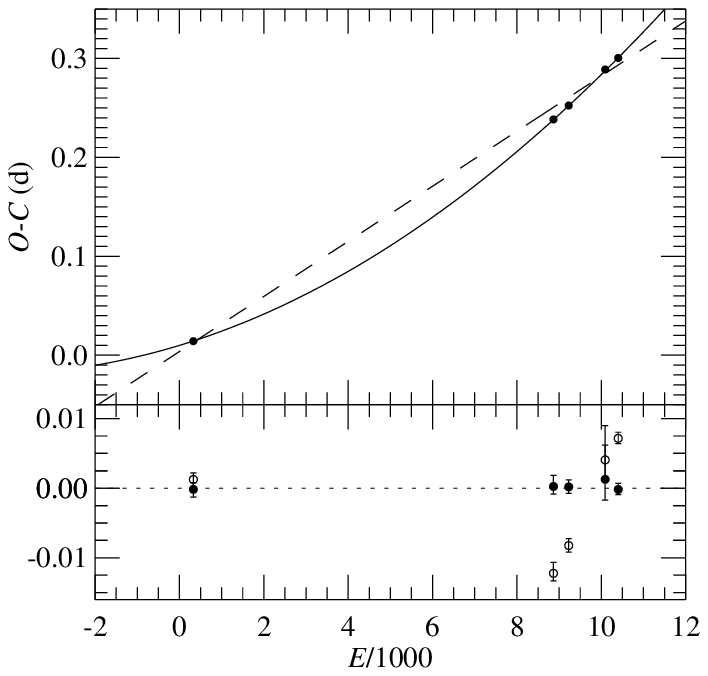}
\caption{The observed-minus-corrected diagram for the primary minimum
  measurements. The solid (dashed) lines indicate the best-fit
  quadratic (linear) models; the residuals relative to these models
  are shown below the \OC\ diagram as filled (open) symbols.} \label{fig:o-c}
%\end{centering}
\end{figure}

We apply the standard \OC\ diagram technique \citep[e.g.,][]{Ste2005}
to assess the evolution of the rotation period. With the reference
epoch defined by \citet{Hes1976} and the rotation period measured by
\citet{Hes1977}, an observed-minus-calculated value
\begin{equation} \label{eqn:o-c}
\OC = \tmin - (2442778\fd819  + 1\fd19801 \cycnum)
\end{equation}
is evaluated for each \tmin\ in Tab.~\ref{tab:minima}; here and in the
table, the cycle number \cycnum\ is the integer that minimizes
$|\OC|$.

Figure~\ref{fig:o-c} plots \OC\ as a function of \cycnum; the error
bars on each point are taken from the corresponding error bounds on
\tmin. Also shown in this \OC\ diagram are fitted linear and quadratic
models of form
\begin{equation} \label{eqn:o-c-linear}
\OCmod = b_{1} + b_{2} \cycnum
\end{equation}
and 
\begin{equation} \label{eqn:o-c-quadratic}
\OCmod = b_{1} + b_{2} \cycnum + b_{3} \cycnum^{2},
\end{equation}
together with the associated fit residuals. The linear model
corresponds to a constant rotation period, while the quadratic one
represents a period that increases linearly with time. The
coefficients $\{b_{j}\}$ are determined by a non-\chisq\ maximum
likelihood estimation, which seeks to minimize the statistic
\begin{equation}
  \OCstat = \sum - \log \PDF [\OC - \OCmod];
\end{equation}
here, the summation is over the five light minima, each with its
respective \OC, \OCmod\ and \PDF(\tmin). For a given model, the
quantity $\exp(-\OCstat)$ is proportional to the likelihood that the
measured \OC\ values could have arisen by chance-fluctuation
departures from the model. We use a downhill simplex algorithm
\citep{Pre1992} to minimize \OCstat, iterated until the maximum
relative deviation between simplex vertices drops below $10^{-4}$.

Table~\ref{tab:o-c} summarizes the coefficients $\{b_{j}\}$ of the
linear and quadratic models. As in \S\ref{ssec:analysis-min}, error
bounds are determined via Monte Carlo simulations. However, rather
than generating synthetic \OC\ data by bootstrapping, we construct
them by perturbing each \tmin\ with random deviates drawn from the
appropriate PDF (cf.~Fig.~\ref{fig:pdf}). From the same simulations we
also determine the distribution of the \OCstat\ statistic, enabling us
to associate the linear and quadratic fits in Fig.~\ref{fig:o-c} with
a likelihood \pnull\ (also specified in Tab.~\ref{tab:o-c}) that the
null hypothesis is true: that is, that the deviation of the \OC\
values from the model arises purely due to chance fluctuations.

\subsection{Period Evolution} \label{ssec:analysis-period}

Table~\ref{tab:o-c} indicates that the null hypothesis for the linear
\OC\ model is \emph{extremely} unlikely ($\pnull = 0.05\%$), allowing
a constant rotation period to be ruled out with a high degree of
confidence. However, the converse is true for the quadratic model,
which fits the data extremely well. Combining its coefficients with
eqns.~(\ref{eqn:o-c}) and~(\ref{eqn:o-c-quadratic}) gives a revised
ephemeris for the primary light minimum of \sorie\ as
\begin{multline} \label{eqn:ephem}
\tmin = 2442778\fd8290^{+0.0012}_{-0.0014} + \mbox{} \\
        1\fd1908229^{+0.0000012}_{-0.0000010} \cycnum + 
1\fd44^{+0.10}_{-0.11} \times 10^{-9} \cycnum^2.
\end{multline}
By taking the derivative with respect to cycle number, the
instantaneous period is found as
\begin{equation} \label{eqn:period}
P = 1\fd1908229^{+0.0000012}_{-0.0000010} +
        2\fd89^{+0.19}_{-0.22} \times 10^{-9} \cycnum,
\end{equation}
which grows linearly at a rate $\dot{P} = 77\,{\rm ms}$ per year. (We
discuss the implicit assumption of \emph{smooth} period growth in
\S\ref{sec:discuss}). At the reference epoch $\cycnum=0$, this period
is rather larger than the $P=1\fd19081 \pm 0\fd00001$ reported by
\citet{Hes1977}; however, these authors' error estimate seems overly
optimistic. \citet{Rei2000} determined a period $P = 1\fd19084 \pm
0\fd00001$ by combining new helium-line data with historical
measurements from \citet{PedTho1977}; their value is in good agreement
with the mean period $\overline{P} = 1\fd190833$ obtained by averaging
eqn.~(\ref{eqn:period}) over the 1976--1998 interval spanned by the
helium data.

\subsection{Systematics} \label{ssec:analysis-sys}

Before discussing the significance of the measured period increase, we
briefly review factors that may have a systematic effect on this
result. As demonstrated by \citet{Tow2008}, the timing of light minima
is sensitive to the optical depth of the magnetospheric plasma
clouds. In principle, the progressive shift in \tmin\ toward later
times (seen in the residuals plot of Fig.~\ref{fig:o-c}) could be
explained by the secular accumulation of plasma in the magnetosphere
\citep[see, e.g.,][]{TowOwo2005}. However, based on the examples given
by \citet[][his section~2.4]{Tow2008}, the observed shift of $\approx
0\fd02$ between the 2004 and 2009 seasons would require a
\emph{factor-six} increase in the optical depth of the
magnetosphere. The minima in Fig.~\ref{fig:minima} clearly do not
exhibit this kind of dramatic change. Indeed, although there
\emph{are} some season-to-season variations in the minima depths (on
the order of $20\,{\rm mmag}$), no monotonic trend is seen; we believe
the variations are probably due to changes in the distribution of
scattered light from \soriab. Accordingly, we rule out the possibility
that the \tmin\ shift is due to plasma accumulation.

Similar reasoning can be addressed to concerns over the change in
filters, from Str\"omgren $u$ in 1977 to Johnson $U$ in the later
observations. In passbands where the magnetosphere is more opaque, the
time of primary light minimum will tend to occur later. This
color-\tmin\ correlation can be clearly seen in the \citet{Hes1977}
observations; in the Str\"omgren $u$ band (falling blueward of the
Balmer jump), the time of primary light minimum is $0\fd007$ later
than in the $v$ band (falling redward of the Balmer jump). Because the
$U$ band straddles the Balmer jump, the expected time lag between $u$
and $U$ should be around half of this, $\sim 0\fd003$. An adjustment
of this order to the 1977 \tmin\ point, to correct for the
color-\tmin\ correlation, has a negligible effect on our results.

A final possible issue comes from the use of parabolae to measure the
primary minimum times. As discussed by \citet{Ste2005}, quadratic
fitting is often eschewed in light-curve analysis on the grounds that
it is unable to adequately model asymmetric light minima. However, if
the shape of the light curve does not vary from cycle to cycle, then
this bias is irrelevant: it matters not that \tmin\ occurs slightly
before or after the precise time of minimum, as long as the lead or
lag remains invariant.

Nevertheless, to explore any bias introduced by the use of quadratic
fitting, we have repeated our analysis using cubic fitting to measure
the \tmin\ values. Three salient points stand out from this
re-analysis: (i) the \chisqr\ values of the cubic fits are not
significantly smaller than those of the quadratic fits
(cf.~Tab.~\ref{tab:minima}); (ii) there's no evidence for a systematic
lag or lead between the quadratic and cubic minima; (iii) there
\emph{is} an obvious difference between the widths of the PDFs, which
are a factor $\sim 2$ broader in the cubic cases than in the quadratic
ones. Thus, we conclude that quadratic minimum fitting does not
introduce any appreciable bias, and moreover is the more robust
approach.

%% Discussion

\section{Discussion} \label{sec:discuss}

The Helium-strong star HD~37776 was found by \citet{Mik2008} to
exhibit a progressive lengthening in its $1\fd5387$ rotation period,
with a characteristic spin-down time $\tspin \equiv P/\dot{P} =
0.25\,{\rm Myr}$. For \sorie\ the absence of photometric data in the
1980s and 1990s means that we cannot empirically differentiate between
steady spin-down and a sequence of abrupt braking episodes. However,
the steady scenario is lent strong support by magnetohydrodynamical
(MHD) simulations of angular momentum loss in magnetically channelled
line-driven winds \citep{udD2009}, which indicate that the lengthening
of rotation periods should be a smooth process. Therefore, the use of
a quadratic ephemeris (cf.~eqn.~\ref{eqn:ephem}) appears justified,
and we derive a characteristic spin-down time $\tspin =
1.34^{+0.10}_{-0.09}\,{\rm Myr}$. This value coincides very well with
the $\tspin = 1.4\,{\rm Myr}$ predicted specifically for \sorie\ by
\citet{udD2009}, from their MHD-calibrated scaling law for spin-down
times. (Such a close agreement is partly fortuitous, given the
uncertainties in stellar and wind parameters). Assuming that \tspin\
has remained constant over the lifetime of the star implies that it
can be no older than $1.16^{+0.09}_{-0.08}\,{\rm Myr}$ (otherwise, it
would at some stage have been rotating faster than the critical rate
$P_{\rm crit} \sim 0\fd5$). This upper limit on the age fits within
the lower portion of the $0.5-8\,{\rm Myr}$ age range estimated for
the $\sigma$ Orionis cluster \citep[][and references
therein]{Cab2007}.

In summary, then, we conclude that the observations are consistent
with \sorie\ undergoing rotational braking due to its magnetized
line-driven wind. This result is significant: although magnetic
rotational braking is inferred from population studies of low-mass
stars \citep[e.g.,][]{DonLan2009}, direct measurement of spin-down in
an individual (non-degenerate) object is noteworthy, and has been
achieved so far for only handful of magnetic B and A stars
\citep[see][]{Mik2009}.

%% Acknowledgements

\acknowledgments

RHDT, DHC and SPO acknowledge support from NASA \emph{Long Term Space
  Astrophysics} grant NNG05GC36G. We thank the referee, Prof. John
Landstreet, for his thoughtful consideration of the paper. This
research has made use of NASA's Astrophysics Data System Bibliographic
Services.

%% References

%\bibliographystyle{aastex}
%\bibliography{spindown}

\begin{thebibliography}{}

\bibitem[\protect\citeauthoryear{{Birney}, {Gonzalez} \& {Oesper}}{{Birney}
  et~al.}{2006}]{Bir2006}
{Birney} D.~S.,  {Gonzalez} G.,    {Oesper} D.,  2006, Observational Astronomy,
  2 edn.
Cambridge University Press, New York

\bibitem[\protect\citeauthoryear{{Caballero}}{{Caballero}}{2007}]{Cab2007}
{Caballero} J.~A.,  2007, \aap, 466, 917

\bibitem[\protect\citeauthoryear{{Donati} \& {Landstreet}}{{Donati} \&
  {Landstreet}}{2009}]{DonLan2009}
{Donati} J.,  {Landstreet} J.~D.,  2009, \araa, 47, 333

\bibitem[\protect\citeauthoryear{{Groote} \& {Hunger}}{{Groote} \&
  {Hunger}}{1982}]{GroHun1982}
{Groote} D.,  {Hunger} K.,  1982, \aap, 116, 64

\bibitem[\protect\citeauthoryear{{Hesser}, {Ugarte} \& {Moreno}}{{Hesser}
  et~al.}{1977}]{Hes1977}
{Hesser} J.~E.,  {Ugarte} P.~P.,    {Moreno} H.,  1977, \apjl, 216, L31

\bibitem[\protect\citeauthoryear{{Hesser}, {Walborn} \& {Ugarte}}{{Hesser}
  et~al.}{1976}]{Hes1976}
{Hesser} J.~E.,  {Walborn} N.~R.,    {Ugarte} P.~P.,  1976, \nat, 262, 116

\bibitem[\protect\citeauthoryear{{Kwee} \& {van Woerden}}{{Kwee} \& {van
  Woerden}}{1956}]{KweVanW1956}
{Kwee} K.~K.,  {van Woerden} H.,  1956, \bain, 12, 327

\bibitem[\protect\citeauthoryear{{Landstreet} \& {Borra}}{{Landstreet} \&
  {Borra}}{1978}]{LanBor1978}
{Landstreet} J.~D.,  {Borra} E.~F.,  1978, \apjl, 224, L5

\bibitem[\protect\citeauthoryear{{Mikul{\'a}{\v s}ek}, {Szasz}, {Krti{\v c}ka},
  {Zverko}, {{\v Z}i{\v z}{\aa}ovsk{\'y}}, {Zejda} \&
  {Gr{\'a}f}}{{Mikul{\'a}{\v s}ek} et~al.}{2009}]{Mik2009}
{Mikul{\'a}{\v s}ek} Z.,  {Szasz} G.,  {Krti{\v c}ka} J.,  {Zverko} J.,  {{\v
  Z}i{\v z}{\aa}ovsk{\'y}} J.,  {Zejda} M.,    {Gr{\'a}f} T.,  2009,
  arXiv:0905.2565

\bibitem[\protect\citeauthoryear{{Mikul{\'a}{\v s}ek et~al.}}{{Mikul{\'a}{\v
  s}ek} et~al.}{2008}]{Mik2008}
{Mikul{\'a}{\v s}ek} Z. et~al.,  2008, \aap, 485, 585

\bibitem[\protect\citeauthoryear{{Oksala} \& {Townsend}}{{Oksala} \&
  {Townsend}}{2007}]{OksTow2007}
{Oksala} M.,  {Townsend} R.~H.~D.,  2007, in {Okazaki} A.~T.,  {Owocki} S.~P.,
   {Stefl} S.,  eds, ASP. Conf. Ser. 361: Active OB-Stars: Laboratories for
  Stellar and Circumstellar Physics p.~476

\bibitem[\protect\citeauthoryear{{Olsen}}{{Olsen}}{1977}]{Ols1977}
{Olsen} E.~H.,  1977, Information Bulletin on Variable Stars, 1332, 1

\bibitem[\protect\citeauthoryear{{Pedersen} \& {Thomsen}}{{Pedersen} \&
  {Thomsen}}{1977}]{PedTho1977}
{Pedersen} H.,  {Thomsen} B.,  1977, \aaps, 30, 11

\bibitem[\protect\citeauthoryear{{Press}, {Teukolsky}, {Vetterling} \&
  {Flannery}}{{Press} et~al.}{1992}]{Pre1992}
{Press} W.~H.,  {Teukolsky} S.~A.,  {Vetterling} W.~T.,    {Flannery} B.~P.,
  1992, {Numerical Recipes in Fortran}, 2 edn.
Cambridge University Press, Cambridge

\bibitem[\protect\citeauthoryear{{Reiners}, {Stahl}, {Wolf}, {Kaufer} \&
  {Rivinius}}{{Reiners} et~al.}{2000}]{Rei2000}
{Reiners} A.,  {Stahl} O.,  {Wolf} B.,  {Kaufer} A.,    {Rivinius} T.,  2000,
  \aap, 363, 585

\bibitem[\protect\citeauthoryear{{Sterken}}{{Sterken}}{2005}]{Ste2005}
{Sterken} C.,  2005, in {Sterken} C.,  ed., ASP Conf. Ser. 335: The Light-Time
  Effect in Astrophysics: Causes and Cures of the O-C Diagram p.~3

\bibitem[\protect\citeauthoryear{{Townsend}}{{Townsend}}{2008}]{Tow2008}
{Townsend} R.~H.~D.,  2008, \mnras, 389, 559

\bibitem[\protect\citeauthoryear{{Townsend} \& {Owocki}}{{Townsend} \&
  {Owocki}}{2005}]{TowOwo2005}
{Townsend} R.~H.~D.,  {Owocki} S.~P.,  2005, \mnras, 357, 251

\bibitem[\protect\citeauthoryear{{ud-Doula}, {Owocki} \& {Townsend}}{{ud-Doula}
  et~al.}{2009}]{udD2009}
{ud-Doula} A.,  {Owocki} S.~P.,    {Townsend} R.~H.~D.,  2009, \mnras, 392,
  1022

\bibitem[\protect\citeauthoryear{{van Dokkum}}{{van Dokkum}}{2001}]{vanD2001}
{van Dokkum} P.~G.,  2001, \pasp, 113, 1420

\end{thebibliography}

\end{document}